\documentclass[pre,showpacs,preprint]{revtex4}
\usepackage{graphicx}

\begin{document}

\title{Non-Abelian self-organized criticality model with one stochastic
site in each avalanche shows multifractal scaling }

\author{Jozef \v{C}ern\'{a}k}

\address{Institute of Physics, P. J. \v{S}af\'{a}rik University in Ko\v{s}ice,
Jesenná 5, Ko\v{s}ice, Slovak Republic}

\begin{abstract}
I have proposed a\textit{ non-Abelian }and stochastic self-organized
criticality model in which each avalanche contains one stochastic
site and all remaining sites in the avalanche are deterministic with
a constant threshold $E_{c}^{I}$. Studies of avalanche structures,
waves and autocorrelations, size moments and probability distribution
functions of avalanche size, for the thresholds $4\leq E_{c}^{I}\leq256$,
were performed. The shell-like avalanche structures, correlated waves
within avalanches, complex size moments and probability distribution
functions show multifractal scaling like the \textit{Abelian} and
deterministic BTW model despite the fact that the model is \textit{non-Abelian
}and stochastic with unbalanced relaxation rules at each stochastic
site. 
\end{abstract}


\maketitle

\section{Introduction}

Bak, Tang, and Wiesenfeld (BTW) \cite{BTW} introduced a concept
of self-organized criticality (SOC) to study dynamical systems with
temporal and spatial degrees of freedom. They designed a simple cellular
automaton with conservative and deterministic relaxation rules to
demonstrate the SOC phenomenon. Manna (M) \cite{Manna} proposed
another conservative SOC model in which stochastic relaxation rules
instead of deterministic rules were defined.

A stable configuration (see below) in the BTW model does not depend
on the order of relaxations so the model is Abelian \cite{Dhar}.
On the other hand, a stable configuration in the M model depends on
the order of relaxations and the model is thus non-Abelian. Dhar \cite{Dhar_abel}
theoretically proved the Abelian property of the M model for the case
when probabilities of many final stable configurations are considered. 

Based on the real-space renormalization group calculations, Pietronero
\emph{at al.} \cite{Pietro} claimed that both deterministic \cite{BTW}
and stochastic \cite{Manna} models belong to the same universality
class, i.e. a small modification in the relaxation rules of the models
cannot change the universality class. It was assumed that both models
show a finite size scaling (FSS) \cite{Chessa}. With FSS the avalanche
size, area, lifetime, and perimeter follow power laws with cutoffs
\cite{Chessa}:

\begin{equation}
P(x)=x^{-\tau_{x}}F(x/L^{D_{x}}),\label{eq:FSS}\end{equation}
 where $P(x)$ is the probability density function of $x$, $F$ is
the cutoff function, and $\tau_{x}$ and $D_{x}$ are the scaling
exponents. The set of scaling exponents $(\tau_{x},\: D_{x})$ defines
the universality class \cite{Chessa}.

Avalanche structure studies \cite{Ben} and numerical simulations
\cite{Lub} showed that the BTW and M models do not belong to the
same universality class. Later, Tebaldi \textit{at al.} \cite{Teb}
found that avalanche size distributions in the BTW model follow multifractal
scaling. They concluded that the avalanche size exponent $\tau_{s}$
(Eq. \ref{eq:FSS}) does not apply to the BTW model. An avalanche
wave decomposition approach \cite{Ivash} was applied \cite{Menech}
to demonstrate the different wave features of the BTW and M models.
Karmakar \emph{at al.} \cite{Karmar} found that models with balanced
relaxation rules are similar to the deterministic BTW model and models
with unbalanced relaxation rules are similar to stochastic M-like
models. These results \cite{Karmar,Teb,Ben,Lub,Menech} support the
idea that the BTW and Manna models are prototypes of different universality
classes.

The BTW and M models are well understood \cite{Dhar,Dhar_abel,Lub,Teb,Ben}.
On the hand, we know very little about the transition from multifractal
to fractal scaling \cite{Cer,Karmar}. I modified the BTW model \cite{BTW}
to study this transition. I allowed stochastic relaxations in one
site in each avalanche. The stochastic site is located in the place
where the avalanche is initiated. Avalanches are randomly initiated
in various lattice sites so the position of stochastic site is changed
at every new avalanche and the stochastic M-like site can visit all
lattice sites for sufficiently large set of avalanches. An avalanche
dynamics study \cite{Menech} is useful to understand the transition
from multifractal to fractal scaling and to test the hypothesis of
precise relaxation balance \cite{Karmar}. The results suggest that
the model cannot belong to either BTW or M universality classes.

In Sec. \ref{sec:model} I introduce a \textit{non-Abelian} and stochastic
model. In Sec. \ref{sec:Results} I apply numerical simulations and
statistical methods to find avalanche structures, autocorrelation
functions, Hurst exponents, avalanche size moments and probability
density functions of avalanche sizes. Section \ref{sec:Discussion}
is devoted to a discussion which is followed by conclusions in Sec.
\ref{sec:Conclusion}.

\section{Stochastic Self-organized criticality model\label{sec:model}}

The stochastic SOC model is defined on a two dimensional (2D) lattice
$L\times L$ where each site $\mathbf{i}$ has assigned two dynamical
variables $E(\mathbf{i})$ and $E_{c}(\mathbf{i})$ \cite{Cer}.
The variable $E(\mathbf{i})$ represents for example, energy and variable
$E_{c}(\mathbf{i})$ represents a threshold at site $\mathbf{i}$.
All thresholds $E_{c}(\mathbf{i})$ are equal to the same value $E_{c}^{I}$
in the interval $4\leq E_{c}^{I}\leq256$. Relaxation rules are undirected,
conservative and deterministic for all sites $\mathbf{i}$ with the
thresholds $E_{c}(\mathbf{i})=E_{c}^{I}$. At each site $\mathbf{i}$
the relaxation rules are precisely balanced \cite{Karmar}. Thus
the sites behave as the BTW sites \cite{BTW} for the threshold $E_{c}^{I}=4$.
In a stable configuration (a stationary state), all sites $\mathbf{i}$
follow $E(\mathbf{i})<E_{c}(\mathbf{i})$. Let us assume that from
a stable configuration we iteratively select $\mathbf{i}$ at random
and increase $E(\mathbf{i})\rightarrow E(\mathbf{i})+1$. If an unstable
configuration is reached, i.e. $E(\mathbf{i})\geq E_{c}^{I}$, then
the site $\mathbf{i}$ is labeled as $\mathbf{i_{\mathrm{M}}}$. The
initial threshold at site $\mathbf{i_{\mathrm{M}}}$, $E_{c}(\mathbf{i_{\mathrm{M}}})=E_{c}^{I}$
is changed to the new value $E_{c}(\mathbf{i_{\mathrm{M}}})=E_{c}^{II}=2$
and stochastic relaxation rules \cite{Manna} are assigned to this
site. In each avalanche only one site $\mathbf{i_{\mathrm{M}}}$ undergoes
stochastic relaxations while all remaining sites relax as undirected,
deterministic and conservative sites. All unstable sites $\mathbf{i}$
(including $\mathbf{i_{\mathrm{M}}}$) where $E(\mathbf{i})\geq E_{c}(\mathbf{i})$
undergo relaxations to reach the stable configuration $E(\mathbf{i})<E_{c}(\mathbf{i})$.
If a stable configuration is reached, then the threshold $E_{c}(\mathbf{i_{\mathrm{M}}})$
at the site $\mathbf{i_{\mathrm{M}}}$ is set to $E_{c}(\mathbf{i_{\mathrm{M}}})=E_{c}^{I}$
and deterministic relaxation rules \cite{BTW} are assigned to the
site $\mathbf{i_{\mathrm{M}}}$. The site $\mathbf{i_{\mathrm{M}}}$
disappears and all sites $\mathbf{i}$ of the lattice are BTW-like
sites. Stable and unstable states are repeated many times. Adding
of energy ($E(\mathbf{i})\rightarrow E(\mathbf{i})+1$) takes place
randomly thus stochastic sites $\mathbf{i_{\mathrm{M}}}$ could visit
all lattice sites. Thus stochastic sites introduce an annealed disorder
in the initial deterministic model \cite{BTW}.

\section{Results\label{sec:Results}}

One stochastic site in each avalanche makes the model stochastic \cite{Biham}. 

\begin{figure}
\includegraphics[width=7cm]{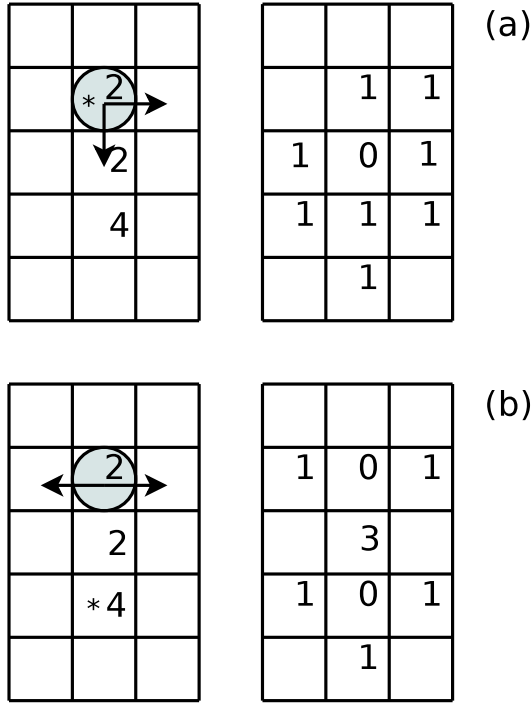}

\caption{\label{fig:non-abelian}(color online) Order of relaxations of the
critical sites shows the\textit{ non-Abelian} property of the OSS
model. The circle denotes a stochastic site and the star ({*}) denotes
a site which relaxes as first. In the initial state there are two
critical sites $E_{c}^{II}=2$ (stochastic) and $E_{c}^{I}=4$. (a)
The stable configuration in the case when the stochastic site ($E_{c}^{II}=2$)
relaxed as first. (b) The stable configuration in the case when the
deterministic site ($E_{c}^{I}=4$) relaxed as first. Arrows show
directions of energy diffusion in the stochastic site. }

\end{figure}

All sites, except one, relax energy as\textit{ Abelian} sites \cite{Dhar_1990,Dhar},
however I classified the model as non-Abelian. To demonstrate the\textit{
non-Abelian} property consider two critical sites \cite{Dhar_1990}
(Fig. \ref{fig:non-abelian}) with thresholds $E_{c}^{II}=2$ (the
stochastic site) and $E_{c}^{I}=4$ (the deterministic site). One
can verify (Fig. \ref{fig:non-abelian}) that the order of relaxations
of unstable sites leads to different final stable configurations i.e.
the OSS model is \textit{non-Abelian}. 

Relaxations in all deterministic sites are precisely balanced \cite{Karmar}.
However, one site in each avalanche shows unbalanced relaxations.
Thus we consider the OSS model for the model with unbalanced relaxation
rules. 

\begin{figure}
\includegraphics[width=7cm]{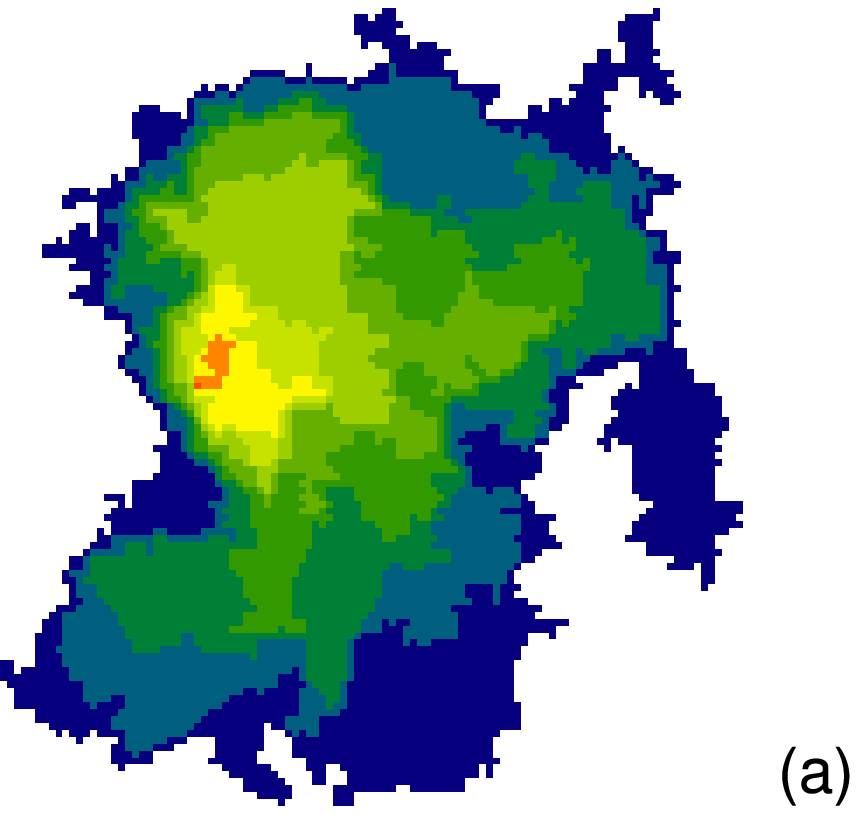}

\includegraphics[width=7cm]{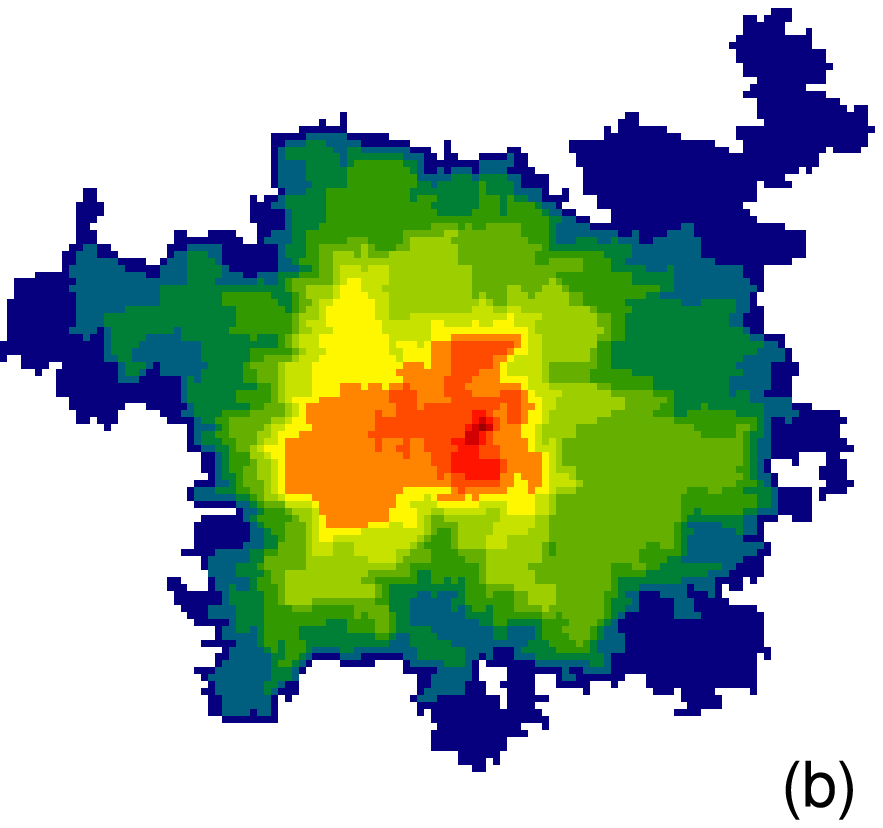}

\caption{\label{fig:aval}(color online) Two types of avalanche structures
on a 2D lattice of size $128\times128$: (a) without holes in the
clusters and (b) with holes in clusters. Lattice sites with the same
numbers of relaxations are shown by the same color (rainbow pseudo-color
coding). Only one site in any avalanche relaxes in a stochastic manner
as a M site (the red area), however all remaining sites relax in a
deterministic manner as BTW sites ($E_{c}^{I}=4$). }

\end{figure}

Analyzing the avalanche structures can provide an important initial
information \cite{Ben,Karmar,Santra} about the nature of the SOC
model. Several avalanches of the OSS model, at the threshold $E_{c}^{I}=4$,
have been decomposed into clusters with equal numbers of relaxations
(Fig. \ref{fig:aval}). Avalanche structures of two types: (i) without
holes Fig. \ref{fig:aval}(a) and (ii) with rare holes in clusters
Fig. \ref{fig:aval}(b) were observed. These structures are similar
to shell-like structures one finds in the BTW model. 

\begin{figure}
\includegraphics[width=7cm]{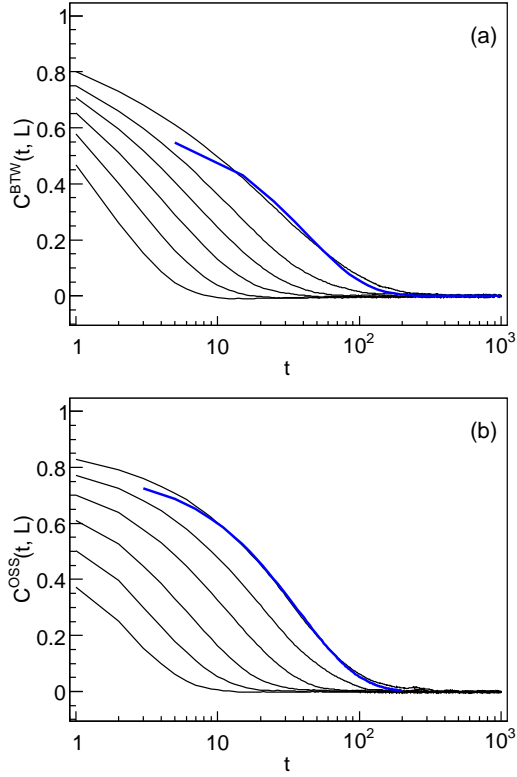}

\caption{\label{fig:Autocorrelation}(color online) Log-lin plots of autocorrelation
functions $C^{BTW}(t,\: L)$ and $C^{OSS}(t,\: L)$ (threshold $E_{c}^{I}=4$)
for lattices: $L=128,\:256,\:512,\:1024$,$\:2048$ and $4096$ (from
left to right). The autocorrelation functions $C^{X}(t,\: L)$ for
given lattice sizes $L$ and for time $t>5$ are approximated by an
exponential function $C^{X}(t,\: L)\sim\exp(-\alpha_{L}t)$ (thick
lines) where decay rates are: (a) $\alpha_{4096}^{BTW}=0.0244$ (BTW
model) and (b) $\alpha_{4096}^{OSS}=0.027$ (OSS model). }

\end{figure}

The study of mathematical SOC models allows avalanches to be decomposed
avalanches into waves \cite{Ivash}. This approach enables the investigation
of correlation of waves within the avalanche \cite{Menech}. I demonstrated
that the OSS model is \textit{non-Abelian,} thus the avalanche waves
could be defined as in the case of the M model \cite{Menech}. On
the other hand, waves initiated by relaxations of the stochastic site
($\mathbf{i_{\mathrm{M}}}$) propagate though the lattice of BTW sites.
Thus waves would be defined as well as waves of the BTW model. An
avalanche of size $s$ is decomposed into $m$ waves with sizes $s_{k}$,
where $s={\textstyle \sum_{k=1}^{m}}s_{k}$. A time-sequence of avalanche
waves $s_{k}$ is used to determine the autocorrelation function \cite{Menech,Stella}

\begin{equation}
C(t,L)=\frac{\langle s_{k+t}s_{k}\rangle_{L}-\langle s_{k}\rangle_{L}^{2}}{\langle s_{k}^{2}\rangle_{L}-\langle s_{k}\rangle_{L}^{2}},\label{eq:autoco}\end{equation}
 where time is $t=1$, $2$, $\ldots$, and the time averages are
taken over $5\times10^{6}$ waves for lattice sizes $L=128$,$\:$$256$,$\:$$512$,$\:$$1024,\:2048$
and $4096$. The autocorrelations $C^{BTW}(t,\: L)$ of the BTW model
(Fig. \ref{fig:Autocorrelation}(a)) and $C^{OSS}(t,\: L)$ of the
OSS model at threshold $E_{c}^{I}=4$ (Fig. \ref{fig:Autocorrelation}(b))
approach zero only for times $t_{max}^{BTW}$ and $t_{max}^{OSS}$
exceeding the maximum number of waves in avalanches. This result is
a consequence of correlated waves in avalanches \cite{Menech}. I
have observed (Fig. \ref{fig:Autocorrelation}) that at a given lattice
size $L$, the time $t_{max}^{OSS}$ is approximately as large as
time $t_{max}^{BTW}$ ($t_{max}^{OSS}\doteq t_{max}^{BTW}$). I note
that avalanche waves in the M model are uncorrelated due to autocorrelation
functions $C^{M}(t,\: L)\doteq0$ for $t\geq1$ \cite{Menech,Karmar}.

The autocorrelations $C^{BTW}(t,\: L)$ (Eq. \ref{eq:autoco}) of
the BTW model were approximated by a power law $f(t)\sim t^{-\tau_{c}}$
and cutoff function $g(t/L^{D_{c}})$ \cite{Menech}:

\begin{equation}
C^{BTW}(t,\: L)=f(t)g(t/L^{D_{c}}).\label{eq:autoco_approx}\end{equation}
 The autocorrelations $C^{BTW}(t,\: L)$ (Fig. \ref{fig:Autocorrelation}(a))
agree well with the previous results \cite{Menech,Karmar}. However,
I have found that the power law approximation $f(t)\sim t^{-\tau_{c}}$
does not approximate the autocorrelation $C^{BTW}(t,\: L)$. I have
verified (Fig. \ref{fig:Autocorrelation}) that the exponential function
$f(t,\: L)\sim\exp(-\alpha_{L}t)$ better approximates not only the
autocorrelation $C^{BTW}(t,\: L)$ but also $C^{OSS}(t,\: L)$. 

\begin{figure}
\includegraphics[width=6.5cm]{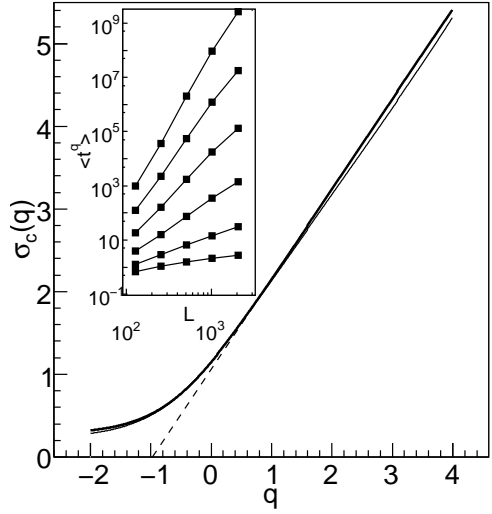}

\caption{\label{fig:timemoments} The plots of $\sigma_{c}(q)$ are approximated
by the linear function $\sigma_{c}(q)=p_{1}q+p_{0}$. For the BTW
model (thin line) the parameters are $p_{1}=1.06\pm0.05,$$\:$$p_{0}=1.06\pm0.05$
and for the OSS model at threshold $E_{c}^{I}=4$ (thick line) the
parameters are $p_{1}=1.09\pm0.05$, and $p_{0}=1.06\pm0.05$ (this
approximation is shown with dashed line). The inset shows log-log
plots of $<t^{q}>$ vs. $L$ for the OSS model and for the exponents
$q=-1,\:0,\:1,\:2,\:3$ and $4$ (from bottom to top). }

\end{figure}

The autocorrelations $C(t,\: L)$ from Eq. \ref{eq:autoco_approx}
are used to compute the time moments \cite{Menech}:

\begin{equation}
\langle t^{q}\rangle_{L}={\displaystyle \sum_{t}C(t,L)t^{q}\sim L^{\sigma_{c}(q)}}.\label{eq:c_moments}\end{equation}
 For lattice sizes $L=128-2048$, threshold $E_{c}^{I}=4$ and for
several values of $q$, the plots of $\log<t^{q}>$ vs. $\log L$
(Fig. \ref{fig:timemoments}, inset) were used to determine the functions
$\sigma_{c}^{BTW}(q)$ and $\sigma_{c}^{OSS}(q)$. The plots $\sigma_{c}^{BTW}(q)$
and $\sigma_{c}^{OSS}(q)$ exhibit linear dependence for $q$ in the
range $1.0\leq q\leq4.0$. From these plots (Fig. \ref{fig:timemoments})
the parameters $D_{c}^{BTW}=1.06\pm0.05$, $D_{c}^{OSS}=1.09\pm0.05$,
$\tau_{c}^{BTW}\doteq0$ and $\tau_{c}^{OSS}\doteq0$ were determined
\cite{Menech}.

\begin{figure}
\includegraphics[width=6.5cm]{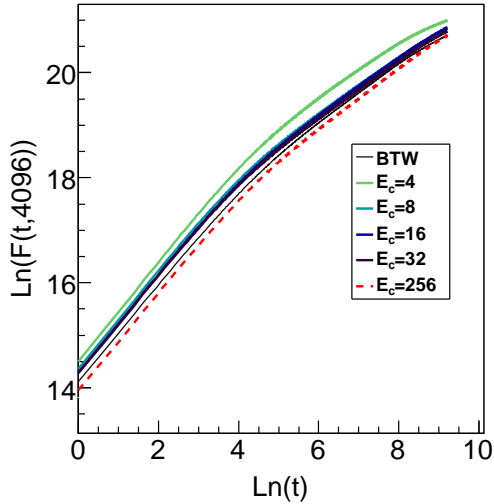}

\caption{\label{fig:Hurst}(color online) The fluctuations $F(t,\: L=4096)$
of the BTW and OSS model with thresholds $4\leq E_{c}^{I}\leq256$.
The Hurst exponents were determined for the threshold $E_{c}^{I}=4$:
$H^{BTW}=0.89\pm0.01$ and $H^{OSS}=0.88\pm0.01$ , the time $1.0<t<50.0$,
$H^{BTW}=0.503\pm0.001$ and $H^{OSS}=0.610\pm0.001$ for the time
$1000.0<t<10000.0$. }

\end{figure}

Stochastic process are often characterized by Hurst exponents \cite{Benoit}.
Fluctuations $F(t,\: L)$ \cite{Menech}:

\begin{equation}
F(t,L)=[\langle\Delta y(t)^{2}\rangle_{L}-\langle\Delta y(t)\rangle_{L}^{2}]^{1/2},\end{equation}
 are used to determine Hurst exponents where $y(t)=\sum_{k=1}^{t}s_{k}$
and $\Delta y(t)=y(k+t)-y(k)$. If fluctuations $F(t,\: L)$ should
scale as $F(t,\:\infty)\sim t^{H}$ then $H$ is the Hurst exponent
\cite{Menech}. Two exponents $H^{BTW}=0.89\pm0.02$ and $H^{BTW}\doteq1/2$
(Fig. \ref{fig:Hurst}) were determined for the BTW model and for
times $t<t_{max}^{BTW}$ and $t>t_{max}^{BTW}$. Similarly, the Hurst
exponents $H^{OSS}=0.88\pm0.02$ and $H^{OSS}=0.610\pm0.001$ were
determined for the OSS model (threshold $E_{c}^{I}=4$) and for times
$t<t_{max}^{OSS}$ and $t>t_{max}^{OSS}$. Fluctuations $F(t,\: L)$
were also determined for the other thresholds $8\leq E_{c}^{I}\leq256$
(Fig. \ref{fig:Hurst}). For all thresholds $4\leq E_{c}^{I}\leq256$
of the OSS model two scaling regions of $F(t,\: L)$ were identified
in contrast to the fluctuation of the M model which exhibits the single
scaling with the Hurst exponent $H^{M}=0.53\pm0.05$ (it is not shown
in Fig. \ref{fig:Hurst}).

\begin{figure}
\includegraphics[width=7cm]{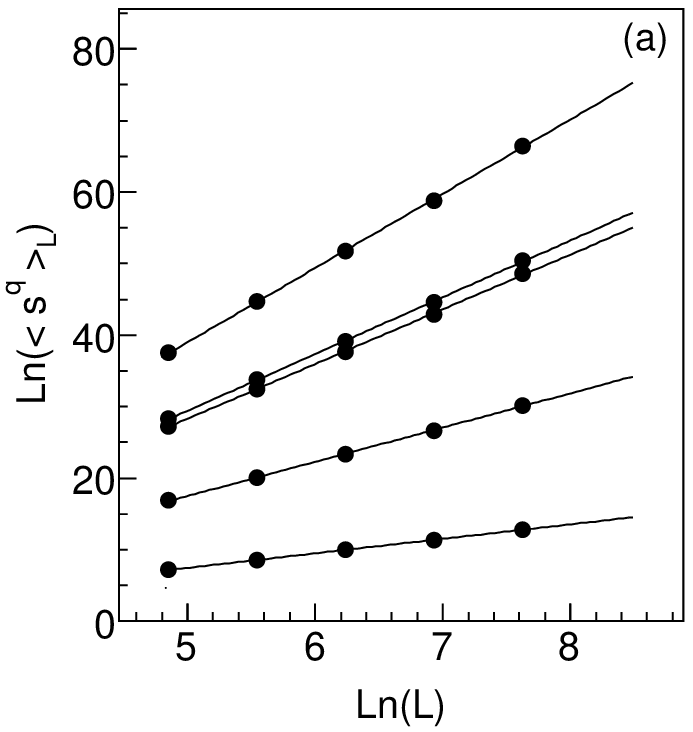}

\includegraphics[width=7cm]{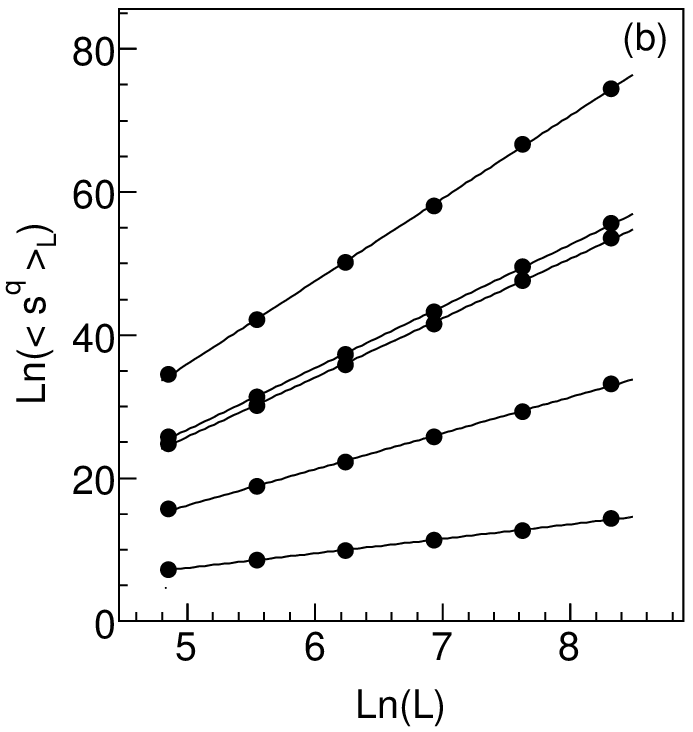}

\caption{\label{fig:Avalanche-size-moments}Avalanche size moments $\langle s^{q}\rangle_{L}$
at lattice size $L$ versus lattice size $L$ display scaling behaviour
for wide range of exponents $q$. The results are shown for (a) the
BTW model ($128\leq L\leq2048$) and (b) the OSS model with threshold
$E_{c}^{I}=4$ ($128\leq L\leq4096$) whereas selected exponents $q$
are $q=1.0,\:2.0,\:3.0,\:3.1$ and $4.0$ (from bottom to top). }

\end{figure}

Moment analysis \cite{Karmar,Teb,Menech} was used to study scaling
properties of both BTW and OSS models. A property $x$ in the FSS
system obeys the scaling given by Eq. \ref{eq:FSS}. The $q$ moments
of $x$ are defined as

\begin{equation}
\langle x^{q}\rangle=\intop_{0}^{x_{max}}x^{q}P(x,L)dx\sim L^{\sigma_{x}(q)},\label{eq:moments}\end{equation}
 where $\sigma_{x}(q)=(q+1-\tau_{x})D_{x}$ and $x_{max}\sim L^{D_{x}}$.
I calculated the moments only for avalanche size $s$. The plots $\log\langle s^{q}\rangle$
versus $\log L$ for approximately five hundred values of the exponent
$q$ were used to determine the functions $\sigma_{s}^{BTW}(q)$ and
$\sigma_{s}^{OSS}(q)$ (Eq. \ref{eq:moments}). These plots scale
precisely for the BTW model (Fig. \ref{fig:Avalanche-size-moments}(a))
and for the OSS model (threshold $E_{c}^{I}=4,$ Fig. \ref{fig:Avalanche-size-moments}(b))
for all exponents $0.0\leq q\leq4.0$ and lattice sizes $128\leq L\leq4096$. 

\begin{figure}
\includegraphics[width=7cm]{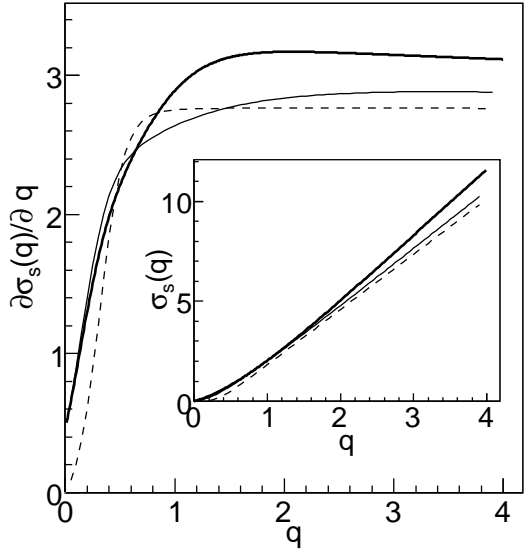}

\caption{\label{fig:dsigma}Plots of $\partial\sigma_{s}(q)/\partial q$ versus
$q$ for the M (dashed curves, $128\leq L\leq2048$), BTW (thin line,
$128\leq L\leq2048$) and OSS models for the thresholds $E_{c}^{I}=4$
(thick solid, $128\leq L\leq4096$). The inset shows plots $\sigma_{s}(q)$
versus $q$ for the M (dashed line, $128\leq L\leq2048$), BTW (thin
line, $128\leq L\leq2048$) and OSS (thick solid, $128\leq L\leq4096$)
models. }

\end{figure}

The results show that $\sigma_{s}^{BTW}(1.0)=2.03$ and $\sigma_{s}^{OSS}(1.0)=2.04$
for threshold $E_{c}^{I}=4$ (Fig. \ref{fig:dsigma}, inset), close
to the expected value $\sigma_{s}^{BTW}(1)\doteq2.0$ \cite{Santra,Karmar}.
The function $\sigma_{s}^{OSS}(q)$ grows faster than functions $\sigma_{s}^{BTW}(q)$
and $\sigma_{s}^{M}(q)$ (Fig. \ref{fig:dsigma}, inset) when the
exponent $q>1.0$ increases. The function $\partial\sigma_{s}^{OSS}(q)/\partial q$
increases with increasing $q>1.0$. At $q=2.07$ it reaches the maximum
$D_{s}^{OSS}(2.07)=3.17\pm0.01$ and for $q>2.07$ is almost constant
or slowly decreases. For $q=4.0$, the capacity dimensions are $D_{s}^{M}(4)=2.76$,
$D_{s}^{BTW}(4)=2.88$ \cite{Cer} and $D_{s}^{OSS}(4)=3.11\pm0.02$.
The functions $\partial\sigma_{s}^{BTW}(q)/\partial q$ and $\partial\sigma_{s}^{OSS}(q)/\partial q$
continuously increase when the exponent $q$ increases in the range
$1.0<q<2.07$. This fact is considered to be a signature of multifractal
scaling \cite{Karmar,Teb}. The capacity dimension of the Manna model
$D_{s}^{M}(q)$ is constant for exponents $q>1.0$: $D_{s}^{M}(q)=2.76\pm0.01$
\cite{Karmar,Cer,Santra}, which is a typical property of the FSS
models \cite{Karmar,Santra}.

Karmakar \textit{et al.} \cite{Karmar} claimed that if local avalanche
dynamics meets criterion of a precise relaxation balance, then the
model must show the same behaviors as the BTW model. If we increase
the threshold $E_{c}^{I}>4$ and we modify the relaxation rules to
meet the criterion of the precise relaxation balance, then the model
with all deterministic sites \cite{Cer2009} has the same properties
as the BTW model. In this model, for thresholds $8\leq E_{c}^{I}\leq256$,
I introduced one stochastic site in each avalanche to compare the
behaviors of the modified SOC model (see Sec. \ref{sec:model}) with
the behaviors of the BTW model. 

\begin{figure}
\includegraphics[width=6.5cm]{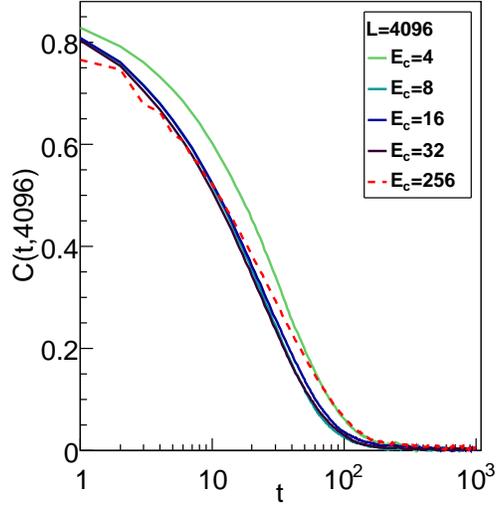}

\caption{\label{fig:Autocorr2}(color online) Autocorrelations $C(t,\: L=4096)$
of the OSS model for thresholds $4\leq E_{c}^{I}\leq256$.}

\end{figure}

Autocorrelations $C(t,\: L)$ Eq. \ref{eq:autoco} were determined
for the OSS model with the thresholds $4\leq E_{c}\leq256$. Avalanches
waves within avalanches are correlated (Fig. \ref{fig:Autocorr2})
for all thresholds because for the time $t<t_{max}$ autocorrelations
$C(t,\: L)$ are greater than $0$ ($C(t,\: L)>0$) and for the time
$t>t_{max}$ the autocorrelations approach the value $C(t,\: L)\doteq0.0$.

\begin{figure}
\includegraphics[width=6.5cm]{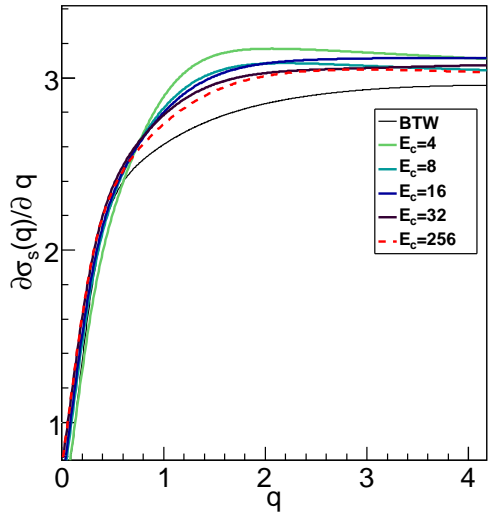}

\caption{\label{fig:dsigma2}(color online) The plots of $\partial\sigma_{s}(q)/\partial q$
of the BTW and OSS ($4\leq E_{c}\leq256$) models ($128\leq L\leq4096$). }

\end{figure}

Functions $\partial\sigma_{s}^{OSS}(q)/\partial q$ of the OSS model
for thresholds $4\leq E_{c}\leq256$ and function $\partial\sigma_{s}^{BTW}(q)/\partial q$
of the BTW model are shown in Fig. \ref{fig:dsigma2}. The results
show that $\partial\sigma_{s}^{OSS}(q)/\partial q>\partial\sigma_{s}^{BTW}(q)/\partial q$
for thresholds $4\leq E_{c}\leq256$ and for exponents $1.0\leq q\leq4.0$.
A difference between functions $\partial\sigma_{s}^{OSS}(q)/\partial q-\partial\sigma_{s}^{BTW}(q)/\partial q$
at given exponent $q$ is observed to be higher than an expected experimental
error \cite{Karmar}. 

\begin{figure}
\includegraphics[width=7cm]{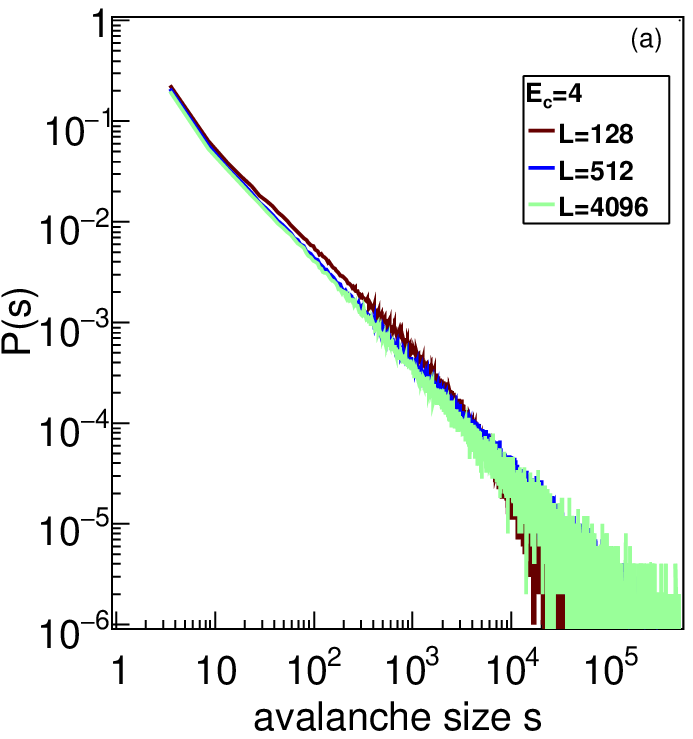}

\includegraphics[width=7cm]{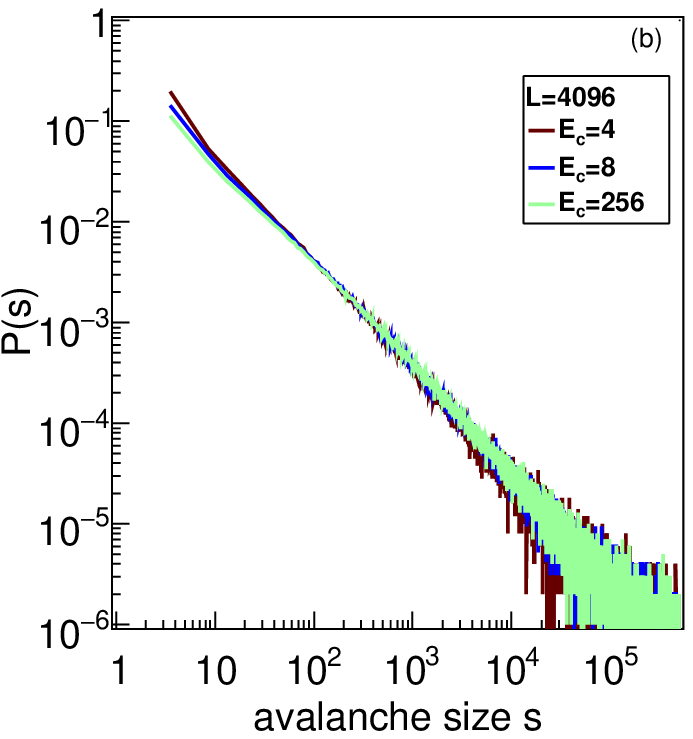}

\caption{\label{fig:pdf}(color online) Probability distribution functions
$P(s)$ of avalanche size $s$ for (a) the constant threshold $E_{c}=4$
and increasing lattice size $L=128,\:512$ and $4096$ and for (b)
the constant lattice size $L=4096$ and increasing thresholds $E_{c}=4,\:8$
and $256$. }

\end{figure}

Probability density functions $P(s)$ of avalanche size $s$ were
determined for different lattice sizes $L=128,\:512$ and $4096$
(Fig. \ref{fig:pdf} (a)) and thresholds $E_{c}^{I}=4,\:8$ and $256$
(Fig. \ref{fig:pdf} (b)) to know the impact of the lattice size $L$
or thresholds $E_{c}^{I}$ on the probability density functions $P(s)$.
The probability density functions $P(s)$ of the OSS model show small
increases of their slopes for avalanches of size $s<10$. 

\section{Discussion\label{sec:Discussion}}

The avalanche structures of the OSS model (Fig. \ref{fig:aval}) are
similar to the shell-like structures of the BTW model\cite{Ben,Teb}.
However, detailed analysis of these structures shows that the structures
do not have the same properties as the structures of the BTW model.
For example, we can see holes inside OSS structures (Fig. \ref{fig:aval}
(b)) which are not possible in the BTW model \cite{Karmar}. These
holes support our classification of the OSS model as a model with
unbalanced relaxation rules \cite{Karmar}.

Autocorrelations $C(t,\: L)$ of the OSS model for thresholds $4\leq E_{c}^{I}\leq256$
(Figs. \ref{fig:Autocorrelation}(b) and \ref{fig:Autocorr2}) are
$C(t,\: L)>0$ for times $t_{max}$. This is a consequence of correlated
avalanche size waves within avalanche \cite{Manna,Stella}. Autocorrelations
of the BTW model are the same as in the paper\cite{Menech}. However,
I cannot confirm that these autocorrelations are approximated by the
power law approximation in Eq. \ref{eq:autoco_approx}. I have observed
that an exponential function $C(t,\: L)\sim\exp(-\alpha t)$ where
$\alpha$ is a decay rate, is a better approximation of the autocorrelations
$C(t,\: L)$ then the power law approximation Eq. \ref{eq:autoco_approx}.
This finding is supported by the fact that time-moment analysis (Fig.
\ref{fig:timemoments}) for the BTW and OSS models leads to the time
exponents $\tau^{BTW}\doteq0$ and $\tau^{OSS}\doteq0$. The reason
for this discrepancy with the previous results \cite{Menech} is
not clear and additional study is necessary. On the other hand, the
maximum number of waves scales as $t_{max}^{BTW}\sim L$ and this
result confirms the previous observation \cite{Menech}. Similarly,
for the OSS model $t_{max}^{OSS}\sim L$ (Fig. \ref{fig:timemoments}).

The fluctuations $F(t,\: L)$ and corresponding Hurst exponents $H_{1}^{BTW}$
and $H_{2}^{BTW}$of the BTW model (Fig. \ref{fig:Hurst}) agree well
with the previous results \cite{Menech,Karmar}. Fluctuations $F(t,\: L)$
of the OSS model for thresholds $4\leq E_{c}^{I}\leq256$ have two
scaling regions (Fig. \ref{fig:Hurst}) and corresponding Hurst exponents
$H_{1}^{OSS}$ and $H_{2}^{OSS}$. An existence of two scaling regions
confirms the correlated avalanche size waves.

The moments of avalanche size Fig. \ref{fig:dsigma} for the BTW and
M models agree well with previous results \cite{Menech,Karmar}.
The plots $\partial\sigma_{s}(q)/\partial q$ of the OSS model in
Figs. \ref{fig:dsigma} and \ref{fig:dsigma2}, for thresholds $4\leq E_{c}^{I}\leq256$
and exponents $q>1.0$, are not constant as in the case of the M model.
The increase of $\partial\sigma_{s}(q)/\partial q$ for $q>1.0$ when
the exponent $q$ increases is considered for a signature of multifractal
scaling \cite{Menech,Teb}. Based on the moment analysis (Figs. \ref{fig:dsigma}
and \ref{fig:dsigma2}) for the thresholds $4\leq E_{c}^{I}\leq256$
and previous conclusions \cite{Menech,Teb,Stella}, I claim that
avalanche size distributions show multifractal scaling.

Probability density functions of avalanche size $P(s)$ of the OSS
model show a moderate increase of small avalanches $s<10$ for thresholds
$4\leq E_{c}^{I}\leq256$ (Fig. \ref{fig:pdf}). I assume that these
changes of probability density functions $P(s)$ do not influence
scaling of avalanche size moments (Fig. \ref{fig:Avalanche-size-moments}).

Holes in some avalanches (Fig. \ref{fig:aval}(b)) are characteristic
for the models with unbalanced relaxation rules which exhibit uncorrelated
avalanche waves \cite{Karmar,Ben}. On the other hand, the existence
of holes in the OSS model is not sign of uncorrelated waves beacuse
avalanche size waves are correlated (see correlations $C^{OSS}(t,L)$
(Fig. \ref{fig:Autocorrelation} (b) and \ref{fig:Autocorr2}) and
fluctuations $F^{OSS}(t,L)$ (Fig. \ref{fig:Hurst})) exhibit two
scaling regions. 

Shell-like avalanche structures \cite{Ben,Karmar}, avalanche wave
correlations $C^{OSS}(t,L)$ \cite{Menech,Karmar}, avalanche wave
fluctuations $F^{OSS}(t,L)$ \cite{Menech} and avalanche size moments
$\sigma_{s}^{OSS}(q)$ \cite{Menech,Karmar} of the OSS model support
the conclusion that the OSS model shows multifractal scaling for thresholds
$4\leq E_{c}^{I}\leq256$. I can reproduce the plots $\partial\sigma_{s}^{BTW}(q)/\partial q$
and $\partial\sigma_{s}^{OSS}(q)/\partial q$ (Fig. \ref{fig:dsigma})
for thresholds $4\leq E_{c}^{I}\leq256$. Comparison of the functions
$\partial\sigma_{s}^{BTW}(q)/\partial q$ with the previous results
of the BTW model and undirected model \cite{Karmar} shows that these
functions collapse to a single function. I demonstrated that the OSS
model shows multifractal scaling but the functions $\partial\sigma_{s}^{OSS}(q)/\partial q$,
for thresholds $4\leq E_{c}^{I}\leq256$, are not identical with the
function $\partial\sigma_{s}^{BTW}(q)/\partial q$ of the BTW model
(Figs. \ref{fig:dsigma}, \ref{fig:dsigma2}). The differences between
functions $\partial\sigma_{s}^{BTW}(q)/\partial q$ and $\partial\sigma_{s}^{OSS}(q)/\partial q$
at given exponent $q$ are larger than the method error \cite{Karmar}.
Based on these evidences, I conclude that OSS and BTW models belong
to the multifractal universality class. However, the models do not
have identical properties despite the fact that in the OSS model only
one site in each avalanche undergoes stochastic relaxations as the
M site.

\section{Conclusion\label{sec:Conclusion}}

The OSS model has been developed to study properties of the inhomogeneous
sand pile model \cite{Cer} at very low densities of M sites \cite{Cer_next}.
Based on traditional classification schemes \cite{Karmar}, one can
expect that the model will belong to the M universality class. However,
I have demonstrated that one stochastic M site in each avalanche is
not enough to change multifractal scaling of the model to the FSS
(Fig. \ref{fig:dsigma}). The OSS model is stochastic, \textit{non-Abelian}
with unbalanced relaxation rules \cite{Karmar}, the classification
schema implies that the model belongs to the M universality class.
However, the OSS models exhibits correlated avalanche waves and multifractal
scaling which is not allowed for the models in the M universality
class. The OSS model exhibits multifractal scaling and the classification
scheme implies that the model belongs to the BTW universality class,
however the model is stochastic, non-Abelian, and has an unbalanced
relaxation rule \cite{Karmar}, thus it cannot belong to the BTW
class. I think that it could be more convenient to consider multifractal
or FSS scaling as a main criterion for model classification to solve
this paradox. Then the models which show multifractal scaling (Fig.
\ref{fig:dsigma}, BTW and OSS models) despite the fact that they
are not identical (see Sec. \ref{sec:Results}) could belong to the
multifractal universality class. Models that show FSS scaling Eq.
\ref{eq:FSS} (M model) could belong to the FSS universality class.
I have analyzed another SOC model where the results support this classification
scheme \cite{Cer2009}.

\

Thanks to Alex Read for reading the manuscript and fort discussion.
Computer simulations were carried out with NorduGrid community resources
and in the KnowARC project. This work was supported by the Slovak
Research and Development Agency under contract No. RP EU-0006-06.

\end{document}